\documentclass[twocolumn,showpacs,preprintnumbers,amsmath,amssymb,latexsym,prd,nofootinbib]{revtex4-1}
\usepackage{latexsym,amssymb,amsfonts,amsmath}
\usepackage{color}
\usepackage[dvips]{graphicx}
\usepackage{slashbox,slashed}
\usepackage{mathrsfs}  
\newcommand{\f}[2]{\frac{#1}{#2}}
\newcommand{\tf}[2]{{\textstyle\f{#1}{#2}}}
\newcommand{\la}{\langle}

\newcommand{\ra}{\rangle}

\newcommand{\nyp}[3]{{\bf #1}, #2 (#3)}
\newcommand{\JHEP}{Jour.\ High Energy Phys.\ }
\renewcommand{\Re}{{\rm Re}\,}

\newcommand{\tr}{{\rm tr}\,}
\newcommand{\diag}{{\rm diag}}
\newcommand{\sgn}{{\rm sgn}\,}

\newcommand{\T}{{\mathscr T}}

\newcommand{\Ha}{{\mathscr H}}
\newcommand{\A}{{\mathscr A}}
\newcommand{\B}{{\mathscr B}}

\newcommand{\Ds}{{\mathscr D}}
\newcommand{\U}{{\mathscr U}}

\newcommand{\Vc}{{\mathcal V}}
\newcommand{\Tc}{{\mathcal T}}

\newcommand{\Hac}{{\mathcal H}}

\newcommand{\Qc}{{\mathcal Q}}

\newcommand{\Dc}{{\mathcal D}}

\begin{document}

\title{Localization and chiral properties near the ordering transition
of an Anderson-like toy model for QCD}

\author{Matteo Giordano}
\email{giordano@bodri.elte.hu}
\affiliation{
Institute for Theoretical Physics, E\"otv\"os University,}
\affiliation{
  MTA-ELTE ``Lend\"ulet'' Lattice Gauge Theory Research Group,\\ P\'azm\'any
  P\'eter s\'et\'any 1/A, H-1117 Budapest, Hungary}

\author{Tam\'as G.\ Kov\'acs}%
\email{kgt@atomki.mta.hu} 
\affiliation{
Institute for Nuclear Research of the Hungarian Academy of Sciences, \\
Bem t\'er 18/c, H-4026 Debrecen, Hungary}%

\author{Ferenc Pittler}%
\email{pittler@hiskp.uni-bonn.de}
\affiliation{
  HISKP(Theory), University of Bonn,\\
  Nussallee 14-16, D-53115 Bonn, Germany}%

\begin{abstract}
  The Dirac operator in finite-temperature QCD is equivalent
  to the Hamiltonian of an unconventional Anderson model, with on-site
  noise provided by the fluctuations of the Polyakov lines. The main
  features of its spectrum and eigenvectors, concerning the density of
  low modes and their localization properties, are qualitatively
  reproduced by a toy-model random Hamiltonian, based on an Ising-type
  spin model mimicking the dynamics of the Polyakov lines. Here we
  study the low modes of this toy model in the vicinity of the ordering
  transition of the spin model, and show that at the critical point
  the spectral density at the origin has a singularity, and the
  localization properties of the lowest modes change. This provides
  further evidence of the close relation between deconfinement, chiral
  transition, and localization of the low modes.
\end{abstract}

\pacs{11.15.Ha,12.38.-t,11.30.Rd,72.15.Rn }

\maketitle

\section{Introduction}
\label{sec:intro}

As is well known, the phase diagram of QCD at zero chemical potential
consists of a low-temperature confining and chirally broken phase, and
a high-temperature deconfined and (approximately) chirally restored
phase. Interestingly enough, the two transitions take place at nearly
the same temperature, or more precisely in the same small temperature
range, as both the deconfining and the chirally restoring transition
are actually steep but nevertheless analytic
crossovers~\cite{Aoki:2005vt,Borsanyi:2010cj}. 
The close connection between the two transitions is even more striking
in certain QCD-like models where they are genuine phase transitions,
like for example SU$(2)$ and SU$(3)$ pure-gauge theories. In this case
lattice calculations show that the deconfinement and the chiral
transitions take place at the very same temperature (of course, within
the inherent numerical uncertainties of lattice
calculations)~\cite{Kogut:1982rt}. The same coincidence of the
transition temperatures has been observed in a model with SU$(3)$
gauge fields and unimproved staggered fermions on 
coarse lattices~\cite{unimproved0,unimproved1,unimproved}. Another
interesting case is that of SU$(3)$ gauge fields with adjoint
fermions: this model is known to possess different deconfinement
($T_d$) and chiral-restoration temperatures ($T_\chi$)~\cite{adjoint},
with $T_d<T_\chi$, but the chiral condensate has a jump exactly at
$T_d$, signaling a first-order chiral phase transition there.
So far, no generally accepted explanation has been provided for the
coincidence of chiral and deconfinement transitions in these models,
and their approximate coincidence in QCD.

In recent years there has been growing evidence that the QCD
finite-temperature transition is accompanied by a change in the
localization properties of the Dirac eigenmodes: while in the
low-temperature phase all the Dirac eigenmodes are delocalized in the
whole volume~\cite{VWrev,deF}, at high temperature the lowest modes
are spatially
localized~\cite{Gockeler:2001hr,GGO,GGO2,KGT,KP,BKS,KP2,feri,crit,Dick:2015twa,GKKP,Giordano:2014qna,  
Cossu:2016scb}. 
This behavior of  the lowest modes is not unique to QCD, and has
been found also in the above-mentioned QCD-like models (i.e., SU$(2)$
and SU$(3)$ pure-gauge theory, and unimproved staggered fermions).
There are indications that the onset of localization takes place
around the same temperature at which QCD becomes deconfined and
chirally restored: this issue was first studied 
by Garc\'ia-Garc\'ia and Osborn in Ref.~\cite{GGO2}. To avoid the
complications related to the crossover nature of the QCD transition, it is
convenient to consider models where the transition is a genuine phase
transition. This was done in Ref.~\cite{GKKP}, which employed the
above-mentioned model with unimproved staggered fermions,
investigating the confining, chiral and localization properties of the
system. In that case, it was found that deconfinement, (approximate)
chiral restoration, and onset of localization take place at the same
value of the gauge coupling, where the system undergoes a first-order
phase transition. These results obviously suggest that localization is
closely related to deconfinement and to the chiral transition.
 
Understanding why the lowest Dirac eigenmodes become localized at the
transition, and how localization affects the corresponding
eigenvalues, might help in shedding some light on the relation between
the deconfining and the chiral transition. As it was suggested
in Ref.~\cite{BKS}, and later elaborated on in more detail in
Refs.~\cite{GKP,GKP2}, localization of the lowest modes is very likely
to be a consequence of deconfinement. More precisely, the ordering of
the Polyakov-line configurations, and the presence therein of
``islands'' of fluctuations away from the ordered value, leads to the
lowest Dirac modes localizing on the ``islands''. In Ref.~\cite{GKP2}
it was suggested that the ordering of the Polyakov lines might also be
responsible for the depletion of the spectral region near the origin,
which in turn leads to a smaller condensate via the Banks-Casher
relation, and so to approximate chiral restoration. 

The argument is most clearly formulated in the Dirac-Anderson approach
of Ref.~\cite{GKP2}. This consists in recasting the
Dirac operator into the Hamiltonian of a three-dimensional system with
internal degrees of freedom, corresponding to color and temporal
momentum. This Hamiltonian contains a diagonal part, related to the
phases of the Polyakov lines, representing a random on-site potential
for the quarks, and an off-diagonal part responsible for their hopping
from site to site, built out of the spatial links on the different
time slices. 
In this framework, the accumulation of eigenmodes near the origin
requires two conditions: sufficiently many sites where
the on-site potential is small, and a sufficiently strong mixing (via
the hopping terms) of the different temporal-momentum components of
the quark wave function. The ordering of the Polyakov lines acts
against both these requirements, by reducing the number of sites where
the potential is small, and localizing them on ``islands'' in a
``sea'' of sites where the potential is large; and by inducing
correlations among spatial links on different time slices, which in
turn makes the mixing of different temporal-momentum components less
effective. This leads to the depletion of the spectral region near the
origin. 

The argument above is based on the results of a detailed numerical
study of a QCD-inspired toy model, constructed in such a way as to
reproduce qualitatively all the important features of the QCD Dirac
spectrum and of the corresponding eigenmodes. In this toy model
the role of the Polyakov lines is played by complex spin
variables, with dynamics determined by an Ising-like model. This spin
model possesses a disordered and an ordered phase, analogous to the
confined and deconfined phases of gauge theories. As 
was shown in Ref.~\cite{GKP2}, the properties of the Dirac spectrum in
the ordered and disordered phases indeed qualitatively match those
found in the deconfined and confined phases of QCD, respectively.
More precisely, deep in the ordered phase the lowest eigenmodes are
localized and the spectral density vanishes near the origin, while in
the disordered phase the lowest eigenmodes are delocalized and the
spectral density is finite near the origin. This makes us confident in
the validity of the mechanism for chiral symmetry restoration
discussed above also in the physically relevant case of QCD.

The magnetization transition of the spin model is expected to be in
the same universality class as that of the 3D Ising model, so one
expects it to be a genuine second-order phase transition. It is thus
worth studying the localization properties of the lowest Dirac
eigenmodes, and the corresponding spectral density near the origin, 
close to the magnetization transition. This is the subject of the
present paper. The purpose is twofold: on the one hand, this model
provides another testing ground for the idea that deconfinement,
chiral transition and localization of the lowest modes are closely
connected. On the other hand, the different order of the transition
with respect to that taking place in the model with unimproved
staggered fermions allows us to study the possible dependence of this
connection on the nature of the transition. 

The paper is organized as follows. In Section \ref{sec:dirac-anderson}
we review the approach to the QCD Dirac spectrum as the spectrum of a
Hamiltonian with noise (``Dirac-Anderson'' approach), considerably
simplifying the formalism of Ref.~\cite{GKP2}. We then briefly recall
the main aspects of the toy model of Ref.~\cite{GKP2}, which we
reformulate equivalently in the new formalism. In Section
\ref{sec:num} we show our numerical results. We first identify
precisely the critical point of the spin model, and then discuss the
localization and chiral properties of our toy model in its
vicinity. Finally, in Section \ref{sec:concl} we report our
conclusions and show our prospects for the future.

\section{The Dirac operator as an Anderson-like Hamiltonian}
\label{sec:dirac-anderson}

In this section we briefly review the derivation of the Dirac-Anderson
form of the staggered Dirac operator, introduced in
Ref.~\cite{GKP2}. We also proceed to simplify the formalism with
respect to the original formulation.

The Dirac-Anderson Hamiltonian is nothing but a suggestive name for
(minus $i$ times) the staggered Dirac operator in the basis of the
eigenvectors of the temporal hopping term. More precisely, denoting it
by $\Hac = -iD_{\rm stag}$, it reads in compact notation 
\begin{equation}
  \label{eq:DAHam}
  \Hac = \eta_4 \Dc  + \f{1}{2i}\sum_{j=1}^3\eta_j\left[
 \Vc_j \Tc_j -\Tc_j^\dag \Vc_j^\dag\right]\,.
\end{equation}
The Dirac-Anderson Hamiltonian, $\Hac_{\vec x a k,\vec y b l}$,
carries space, color and temporal-momentum indices, 
for $\vec x,\vec y \in \mathbb{Z}_{L}^3$, $a,b = 1,\ldots, N_c$, and
$k,l = 0,\ldots,N_T-1$. Here $\mathbb{Z}_{L}^3=\{\vec x\, |\, 0\le
x_i\le L-1 \}$, and $L$ and $N_T$ are the spatial and temporal
extension of the lattice, which have to be even integer
numbers. Periodic boundary conditions in the spatial directions are
understood.\footnote{
Antiperiodic boundary conditions in the temporal direction are of
course understood in the original four-dimensional staggered operator,
and they reflect in the form of the effective Matsubara frequencies
given below in Eq.~\eqref{eq:matsu}. However, since the Dirac-Anderson
Hamiltonian is a three-dimensional Hamiltonian, there are no temporal
boundary conditions to be imposed on the fermions.} 
In Eq.~\eqref{eq:DAHam}, $\Dc$ is the
diagonal matrix consisting of the ``unperturbed'' eigenvalues of the
temporal hopping term, $\Vc_j$ come from the spatial hoppings, and
$\Tc_j$ is the translation operator in direction $j$,
\begin{equation}
  \label{eq:notation}
  \begin{aligned}
    (\Dc)_{\vec x a k,\vec y b l} &=  
\delta_{\vec x \vec y}\delta_{ab}\delta_{kl} \sin\omega_{ak}(\vec x)\,,\\
(\Vc_j)_{\vec x a k,\vec y b l} &=  \delta_{\vec x \vec y}(V_{+j}(\vec
x))_{ak,bl}\,,\\ 
(\Tc_j )_{\vec x a k,\vec y b l} &=\delta_{\vec x+\hat\jmath, \vec
  y}\delta_{ab}\delta_{kl}\,,
  \end{aligned}
\end{equation}
and moreover $\eta_\mu=(-1)^{\sum_{\nu<\mu} x_\nu}$ are the usual staggered
phases. Let us explain the notation in detail.
The effective Matsubara frequencies $\omega_{ak}(\vec x)$ are given by
\begin{equation}
  \label{eq:matsu}
  \omega_{ak}(\vec x) = \f{\pi + \phi_a(\vec x) + 2\pi k}{N_T}\,,
\end{equation}
with $\phi_a(\vec x)$ being the phases of the Polyakov line $P(\vec
x)=\prod_{t=0}^{N_T-1}U_4(t,\vec x)$. The following convention is
chosen for the Polyakov-line phases: $\phi_a(\vec x)\in [-\pi,\pi)$
for $a=1,\ldots,N_c-1$, and
$\sum_a \phi_a(\vec x) =0$.\footnote{A redefinition modulo $2\pi$
  corresponds simply to a unitary transformation of the
  Hamiltonian~\cite{GKP2}.} 
The spatial hoppings read
\begin{equation}
  \label{eq:off_diag2}
  \begin{aligned}
 ( V_{+j}(\vec x) )_{ak,bl} = \f{1}{N_T}\sum_{t=0}^{N_T-1}& e^{i\f{2\pi
      t}{N_T}(l-k)}
  e^{i\f{t}{N_T}(\phi_b(\vec x +\hat\jmath)-\phi_a(\vec x))}\\ &\times
 ( U^{\rm (td)}_j(t,\vec x))_{ab}\,,
  \end{aligned}
\end{equation}
where $U^{\rm (td)}_j(t,\vec x)$ is the gauge link corresponding to
the lattice link $(t,\vec x)\to (t,\vec x+\hat\jmath)$ in the temporal
diagonal gauge (or Polyakov gauge), $U^{\rm (td)}_4(t,\vec
x)=\mathbf{1}$ for $0\le t < 
N_T-1$ and $U^{\rm (td)}_4(N_T-1,\vec x)=\diag(e^{i\phi_a(\vec x)})$. 
One can show that $V_{+j}(\vec x)$ is a unitary matrix in color and
temporal-momentum space.

The expression Eq.~\eqref{eq:DAHam} is obviously fully equivalent to
the staggered Dirac operator. Moreover, its structure is reminiscent
of a 3D Anderson Hamiltonian with internal degrees of freedom 
corresponding to color and temporal momentum, and with antisymmetric
rather than symmetric hopping term. The diagonal noise is  
provided by the phases of the Polyakov lines. The off-diagonal
noise present in the hopping terms comes both from the spatial links and
from the Polyakov-line phases. The amount of disorder is controlled by
the size of the fluctuations of the Polyakov lines and of the spatial
links, and therefore by the temperature of the system (as well as the
lattice spacing). 

Differently from the usual Anderson models, the strength of the
disorder is fixed, since the absolute value of the diagonal terms is
bounded by 1, and since the hopping terms are unitary matrices. What is
different on the two sides of the deconfinement transition is the
distribution of the diagonal terms, and the matrix structure of the
hoppings. Indeed, at high temperature the ordering of the Polyakov 
line leads to the enhancement of diagonal terms corresponding to the
trivial phase $\phi_a(\vec x)=0$, which form a ``sea'' of large (i.e.,
close to 1) unperturbed eigenvalues. Fluctuations away from the
trivial phase form localized ``islands'' of smaller unperturbed
eigenvalues. Moreover, the ordering of the Polyakov lines leads to
strong correlations among spatial links on different
time slices. These correlations tend to reduce the off-diagonal
entries of the hopping term in temporal-momentum space in the ``sea''
region, thus approximately decoupling the different temporal-momentum
components of the quark wave function.  At low temperatures, on the
other hand, correlations across time slices are weaker, and the
different temporal-momentum components of the quark wave function mix 
effectively. 

\subsection{Simplifications of the Dirac-Anderson Hamiltonian}
\label{sec:simpli}

We now discuss a few convenient simplifications of the Dirac-Anderson
Hamiltonian, Eq.~\eqref{eq:DAHam}. First of all, by making a suitable
gauge transformation we will disentangle the two sources of noise,
i.e., we will make the hopping terms independent of the Polyakov-line
phases. 
Let us define
\begin{equation}
  \label{eq:rootofPL}
  W(\vec x) = \diag\left(e^{i\f{\phi_a(\vec x)}{N_T}}\right)\,,
\end{equation}
which satisfies $W(\vec x)^{N_T}=P(\vec x)$, and moreover is easily
seen to be unitary and unimodular, thanks to our choice of convention for
the phases of the Polyakov lines. Eq.~\eqref{eq:off_diag2} can then be 
recast as 
\begin{equation}
  \label{eq:off_diag4}
  V_{+j}(\vec x) = \f{1}{N_T}\sum_{t=0}^{N_T-1} 
e^{i\f{2\pi t}{N_T}(l-k)}
\hat U_j(t,\vec x)\,,
\end{equation}
where
\begin{equation}
  \label{eq:gauge_transf}
\hat U_j(t,\vec x) =[W(\vec x)^\dag]^t U^{\rm (td)}_j(t,\vec x)
[W(\vec x+\hat\jmath)]^t\,.   
\end{equation}
Since $W(\vec x)\in {\rm SU}(N_c)$, Eq.~\eqref{eq:gauge_transf} is just a
gauge transformation, that leads to the 
``uniform diagonal'' gauge: since $\hat U_4(t,\vec x) = [W(\vec x)^\dag]^t
W(\vec x)^{t+1}=W(\vec x)$, one has that the temporal links are
constant and diagonal. 
For future reference, we notice that in this gauge the contribution of
time-space plaquettes to the  Wilson action, which in the temporal
diagonal gauge is proportional to 
\begin{equation}
  \label{eq:action_new}
  \begin{aligned}
\Delta S_{ts}    &= 
\sum_{j,\vec x}\sum_{t=0}^{N_T-2} \Re\tr[  U^{\rm (td)}_j(t,\vec x) U^{\rm
  (td)}_j(t+1,\vec x)^\dag] \\ 
&\phantom{=}+ \sum_{j,\vec x}\Re\tr[  U^{\rm (td)}_j(N_T-1,\vec x) 
P(\vec x + \hat\jmath) \\ &\phantom{=+ \sum_{j,\vec x}\Re\tr[}
\times U^{\rm
  (td)}_j(0,\vec x)^\dag P(\vec x)^\dag]\,,
\end{aligned}
\end{equation}
becomes
\begin{equation}
  \label{eq:action_new_ter}
  \begin{aligned}
  \Delta S_{ts}    =
\sum_{j,\vec x}\sum_{t=0}^{N_T-1} \Re \tr [& \hat U_j(t,\vec x) 
W(\vec x+\hat\jmath)
\\ &\times \hat U_j(t+1,\vec x)^\dag W(\vec x)^\dag ] \,.
  \end{aligned}
\end{equation}
The form of the space-space plaquettes is unaffected by the gauge
transformation, and so 
\begin{equation}
  \label{eq:action_new_ss}
  \begin{aligned}
\Delta S_{ss}    &= 
\sum_{j< j',\vec x,t}\Re\tr[  U^{\rm (td)}_j(t,\vec x) U^{\rm (td)}_{j'}(t,\vec x+ \hat \jmath)
\\ & \phantom{=\sum_{j< j',\vec x,t}\Re\tr[}\times U^{\rm (td)}_j(t,\vec x+\hat\jmath\,' )^\dag U^{\rm
  (td)}_{j'}(t,\vec x)^\dag]\\
&= \sum_{j< j',\vec x,t}\Re\tr[  \hat U_j(t,\vec x) \hat U_{j'}(t,\vec
x+ \hat \jmath)
\\ & \phantom{= \sum_{j< j',\vec x,t}\Re\tr[}\times\hat U_j(t,\vec x+\hat\jmath\,' )^\dag \hat U_{j'}(t,\vec x)^\dag]\,.
\end{aligned}
\end{equation}
The second simplification is obtained by using the following property
of the diagonal entries,  
\begin{equation}
  \label{eq:sym_diag}
  \sin\omega_{a\bar{k}}(\vec x) = -\sin\omega_{ak}(\vec
  x)\,,~~\bar{k}\equiv\left(k+\f{N_T}{2}\right)_{N_T}\,, 
\end{equation}
where $\left(a+b\right)_{N_T} \equiv a+b \mod N_T$, and the cyclicity
of $V_{+j}$, in particular the property
\begin{equation}
  \label{eq:cycl_hop}
 (V_{+j}(\vec x))_{ak,bl}=(V_{+j}(\vec x))_{a\bar{k},b\bar{l}} \,.
\end{equation}
This allows us 
to organize the matrices $\Dc$ and $\Vc_j$  in blocks of size
$\f{N_T}{2}\times\f{N_T}{2}$. Explicitly, we can write
\begin{equation}
  \label{eq:sym_diag2}
  \begin{aligned}
    \Dc &=
    \begin{pmatrix}
      \Ds &  \mathbf{0}\\ 
      \mathbf{0} & -\Ds
    \end{pmatrix}=\Ds \Sigma_3\,, \\ 
    \Vc_j &=
    \begin{pmatrix}
      \A_j & \B_j \\ 
      \B_j & \A_j  
    \end{pmatrix} = \A_j + \B_j\Sigma_1\,,
  \end{aligned}
\end{equation}
where
\begin{equation}
  \label{eq:off_diag}
  \begin{aligned}
  (\Ds)_{\vec x a k,\vec y b l} &=  
\delta_{\vec x \vec y}\delta_{ab}\delta_{kl} \sin\omega_{ak}(\vec
x)\,,
\\
(\A_j)_{\vec x a k,\vec y b l}&=(\Vc_j)_{\vec x a k,\vec y b l}\,,\\
(\B_j)_{\vec x a k,\vec y b l}&=(\Vc_j)_{\vec x a k,\vec y b
  \bar{l}}\,,
  \end{aligned}
\end{equation}
with $k,l = 0,\ldots,\tf{N_T}{2}-1$,
and where $\Sigma_i= \sigma_i\otimes\mathbf{1}_{\f{N_T}{2}}$, i.e.,
\begin{equation}
  \label{eq:sigma}
  \Sigma_1=
\begin{pmatrix}
  \mathbf{0} & \mathbf{1}\\
\mathbf{1} & \mathbf{0}
\end{pmatrix}\,, \quad
\Sigma_2=
\begin{pmatrix}
  \mathbf{0} & -i\mathbf{1}\\
i\mathbf{1} & \mathbf{0}
\end{pmatrix}\,, \quad
\Sigma_3=
\begin{pmatrix}
  \mathbf{1} & \mathbf{0}\\
\mathbf{0} & -\mathbf{1}
\end{pmatrix}
\,.
\end{equation}
For future utility we also define
\begin{equation}
  \label{eq:T_smaller}
  (\T_j )_{\vec x a k,\vec y b l} =\delta_{\vec x+\hat\jmath, \vec
  y}\delta_{ab}\delta_{kl}\,,\quad
k,l = 0,\ldots,\tf{N_T}{2}-1\,.
\end{equation}
We now make use of the block structure of the Dirac-Anderson
Hamiltonian [see Eq.~\eqref{eq:sym_diag2}], and of the fact that it
anticommutes with the unitary matrix $  \Qc = \eta_4
\Sigma_1$,\footnote{This is the analogue of the well-known
  anticommutation relation of the staggered operator with
  $\eta_5\equiv (-1)^{\sum_\nu x_\nu}$.} to simplify the study of the
eigenvalue problem. The eigenvectors of $\Qc$ are of the form
\begin{equation}
  \label{eq:Q}
  \psi_\pm[\varphi] = \f{1}{\sqrt{2}}
  \begin{pmatrix}
    \varphi \\ \pm\eta_4 \varphi
  \end{pmatrix}\,,
\qquad \Qc\psi_\pm[\varphi] = \pm \psi_\pm[\varphi]\,,
\end{equation}
where $\varphi$ are $\f{N_T}{2}$-dimensional. One can easily show that 
\begin{equation}
  \label{eq:Q2}
  \begin{aligned}
  \Sigma_1 \psi_\pm[\varphi] &= \pm\eta_4 \psi_\pm[\varphi]\,,\qquad
\Sigma_3 \psi_\pm[\varphi] = \psi_\mp[\varphi]\,,\\
\Tc_j\psi_\pm[\varphi] &= \psi_\mp[\T_j\varphi]\,.
  \end{aligned}
\end{equation}
Making use of this we find
\begin{equation}
  \label{eq:H_1}
  \begin{aligned}
  \Hac\psi_\pm[\varphi] &= \psi_\mp[\Ha_\pm \varphi]     \,,\\
  \Ha_\pm &= \eta_4 \Ds + \f{1}{2i}\sum_j \eta_j
\left[\U_j^\mp\T_j
-\T_j^\dag\U_j^\pm{}^\dag\right]\,,
  \end{aligned}
\end{equation}
where the matrices
\begin{equation}
  \label{eq:Udef}
\U_j^\pm \equiv \A_j \pm \eta_4\B_j
\end{equation}
are unitary, as a consequence of the unitarity of $\Vc_j$.
One can also prove that $\det \Vc_j = \det(\A_j +\B_j) \det(\A_j -\B_j) 
= \det\U_j^+\det\U_j^-$. From the orthogonality of $\psi_+$ and $\psi_-$
it follows that
\begin{equation}
  \label{eq:sc_prod2}
  (\psi_{s_1}[\varphi_{k}],\Hac\psi_{s_2}[\varphi_{k'}])
=\delta_{s_1, - s_2}(\varphi_{k},\Ha_{s_2}\varphi_{k'})\,,
\quad s_{1,2}=\pm\,,
\end{equation}
i.e., in the basis $\psi_{\pm}[\varphi_{k}]$, with $\{\varphi_{k}\}$ a
basis of the $V\cdot \f{N_T}{2}\cdot N_c$-dimensional space, one finds
\begin{equation}
  \label{eq:DAHam4}
  [\Hac] =
    \begin{pmatrix}
      0 &  \Ha_-\\ \Ha_+ & 0
    \end{pmatrix}\,.
\end{equation}
In order to determine the spectrum of $\Hac$, it is convenient to first
diagonalize $\Hac^2$,
\begin{equation}
  \label{eq:DAH2am4}
  [\Hac^2] =
    \begin{pmatrix}
       \Ha_-\Ha_+ & 0 \\ 0 & \Ha_+\Ha_-
    \end{pmatrix}=
    \begin{pmatrix}
       \Ha_+^\dag \Ha_+ & 0 \\ 0 & \Ha_+\Ha_+^\dag
    \end{pmatrix}\,.
\end{equation}
If $\Ha_+^\dag \Ha_+\varphi_{\lambda^2}=\lambda^2\varphi_{\lambda^2}$,
with  $\lambda\neq 0$, then we have
$(\Ha_+ \Ha_+^\dag) \Ha_+\varphi_{\lambda^2}=\Ha_+(\Ha_+^\dag
\Ha_+)\varphi_{\lambda^2}=\lambda^2\Ha_+\varphi_{\lambda^2}$, so that 
$\f{\Ha_+}{\sqrt{\lambda^2}}\varphi_{\lambda^2}$ is a normalized
  eigenvector of $\Ha_+ \Ha_+^\dag$ with eigenvalue $\lambda^2$ if 
$\varphi_{\lambda^2}$ is a normalized
  eigenvector of $\Ha_+^\dag \Ha_+$. In conclusion, the eigenvectors of
  $\Hac^2$ are of the form $\psi_+[\varphi_{\lambda^2}]$ and
  $\psi_-[\f{\Ha_+}{\sqrt{\lambda^2}}\varphi_{\lambda^2}]$. 

In this paper we are interested in the localization properties of the
eigenmodes. As discussed in Ref.~\cite{GKP2}, a convenient measure of
localization is provided by the participation ratio ${\rm
  PR}={\rm IPR}^{-1}/V$, 
where $V=L^3$ is the lattice volume and ${\rm IPR}$ is the inverse
participation ratio, defined as 
\begin{equation}
  \label{eq:genIPR}
  {\rm IPR} = \sum_{\vec x} \left(\sum_{a,k} |\psi_{ak}(\vec
    x)|^2\right)^2\,.
\end{equation}
With this definition, the knowledge of $\varphi_{\lambda^2}$ is
sufficient to determine the ${\rm IPR}$: indeed,
\begin{equation}
  \label{eq:genIPR2}
  \begin{aligned}
  {\rm IPR}&= \sum_{\vec x} \left(\sum_{a,k} |\psi_{\pm ak}[\varphi(\vec
    x)]|^2\right)^2
\\  &=\sum_{\vec x} \left(\sum_{a,\,0\le k< \f{N_T}{2}} |\varphi_{ak}(\vec
    x)|^2\right)^2 \,.
  \end{aligned}
\end{equation}
For our purposes the problem is thus reduced to a 
$V\cdot \f{N_T}{2}\cdot N_c$-dimensional one. This reduction is the
analogue, in the present basis, of the well-known reduction of $D_{\rm
  stag}^2$ to the sum of two operators, each of which connects only even or
odd sites, in the usual (coordinate) basis.

\subsection{Dirac-Anderson Hamiltonian for $N_T=N_c=2$}
\label{sec:ntnc2}

In the case $N_T=N_c=2$ the problem simplifies considerably. In this
case $\f{N_T}{2}=1$, so a single temporal-momentum component has to be
considered, and $\U^\pm_j$ have the same dimensionality as $\hat
U_j$. We have 
\begin{equation}
  \label{eq:nt2nc2_1}
  \Ds = \cos\f{\phi}{2}\mathbf{1}_{\rm c}\,,
\end{equation}
where $\cos\f{\phi}{2}=\diag(\cos\f{\phi_{\vec x}}{2})$ is a diagonal
matrix in position space, and $\mathbf{1}_{\rm c}$ is the identity in
color space. Moreover,
\begin{equation}
  \label{eq:nt2nc2_2}
  \begin{aligned}
    \A_j(\vec x) &= \f{1}{2}\left(\hat U_j(0,\vec x) +\hat U_j(1,\vec
      x) \right)\,, \\
    \B_j(\vec x) &= \f{1}{2}\left(\hat U_j(0,\vec x) -\hat U_j(1,\vec
      x) \right)\,, 
  \end{aligned}
\end{equation}
and so
\begin{equation}
  \label{eq:nt2nc2_3}
    \U_j^\pm(\vec x) = \delta_\pm(\vec x) \hat U_j(0,\vec x) +
    \delta_\mp(\vec x)\hat 
    U_j(1,\vec x)  \,,
\end{equation}
where $\delta_\pm$ are the projectors on the even and the odd
sublattices, 
\begin{equation}
  \label{eq:eo_proj}
   \delta_\pm(\vec x) = \f{1\pm \eta_4(\vec x)}{2}\,,\qquad
\delta_\pm^2 = \delta_\pm\,, \quad \delta_\pm\delta_\mp = 0\,.
\end{equation}
Inverting these relations we find
\begin{equation}
  \label{eq:nt2nc2_4}
  \begin{aligned}
\hat U_j(0,\vec x)&=  \delta_+(\vec x)\U_j^+(\vec x) +  \delta_-(\vec
x)\U_j^-(\vec x) \,,\\
\hat U_j(1,\vec x)&=  \delta_+(\vec x)\U_j^-(\vec x) +  \delta_-(\vec
x)\U_j^+(\vec x)\,.     
  \end{aligned}
\end{equation}
Notice that changing integration variables to $\U_j^\pm$ leaves the
link integration measure unchanged. 
Let us work out in detail the contribution $\Delta S_{ts}$ to the
action. Since
\begin{equation}
  \label{eq:nt2nc2_5}
  \begin{aligned}
  W(\vec x) &= \diag(e^{i\f{\phi(\vec x)}{2}},e^{-i\f{\phi(\vec x)}{2}}) 
\\ &= \cos\f{\phi(\vec x)}{2}\mathbf{1}_{\rm c} + i\sin \f{\phi(\vec
  x)}{2}(\sigma_3)_{\rm c}\,, 
  \end{aligned}
\end{equation}
after simple algebra one finds
\begin{equation}
  \label{eq:nt2nc2_6}
  \begin{aligned}
    \Delta S_{ts}     &=
2\sum_{j,\vec x} 
\cos\f{\phi(\vec x)}{2}\cos\f{\phi(\vec x+\hat\jmath)}{2}
\Re\tr[  \U_j^+(\vec x)  \U_j^-(\vec x)^\dag] \\ &+
\sin\f{\phi(\vec x)}{2}\sin\f{\phi(\vec x+\hat\jmath)}{2}
\Re\tr [ \U_j^+(\vec x)\sigma_3  \U_j^-(\vec x)^\dag \sigma_3]
 \,.
  \end{aligned}
\end{equation}
As for the Hamiltonian, it is entirely determined by 
\begin{equation}
  \label{eq:nt2nc2_7}
    \Ha_\pm = \eta_4 \cos\f{\phi}{2} + \f{1}{2i}\sum_{j=1}^3 \eta_j
\left[\U_j^\mp\T_j
-\T_j^\dag\U_j^\pm{}^\dag\right]\,.
\end{equation}

\subsection{Toy model}
\label{sec:toy}

The toy model of Ref.~\cite{GKP2} consists simply in replacing the
Polyakov-line phases and spatial links in the various terms appearing
in Eq.~\eqref{eq:DAHam} with suitable toy-model variables, and in
choosing appropriate dynamics for these variables, intended to
mimic that of the corresponding variables in QCD. In particular, the
(diagonal) Polyakov lines $e^{i\phi_a(\vec x)}$ are replaced by complex
spin variables $s^a_{\vec x}=e^{i\phi^a_{\vec x}}$, with dynamics
governed by a suitable spin model. 
The only thing changing for the spatial links is the dynamics,
which is still determined by a Wilson-like action (in the temporal
diagonal gauge), obtained by dropping the contributions from spatial
plaquettes, replacing the Polyakov lines with the diagonal matrices
$\diag(s^a_{\vec x})$, and omitting the backreaction of the gauge
links on the spins, i.e., treating the spins as external fields for
the gauge links. The backreaction of fermions in the partition
function is also omitted, i.e., the fermion determinant is dropped.   

The simplifications of the Dirac-Anderson Hamiltonian discussed
previously translate directly into simplifications for the toy
model. Indeed, such simplifications are obtained by means of a  
gauge transformation for the link variables and of a change of basis
for the Hamiltonian. In both cases, they amount to a unitary
transformation of the Hamiltonian, which therefore leaves the spectrum
unchanged. Moreover, since these transformations are local in space,
they do not alter the localization properties of the eigenmodes. The
toy model obtained by making the substitutions discussed in the
previous paragraph in the Hamiltonian $[\Hac]$, Eq.~\eqref{eq:DAHam4}, 
is thus unitarily equivalent to the one obtained by making the same
substitutions in Eq.~\eqref{eq:DAHam}. In the case $N_c=N_T=2$, which
is the one studied numerically in Ref.~\cite{GKP2}, one can also
make a change of variables for the links, as described in
Eq.~\eqref{eq:nt2nc2_4}, leading to further simplifications.

All in all, the toy model for $N_c=N_T=2$ of Ref.~\cite{GKP2} can be
equivalently formulated as follows. The toy model Hamiltonian reads
\begin{equation}
  \label{eq:toy1}
  \begin{aligned}
  \Hac^{\rm toy} &=
    \begin{pmatrix}
      0 &  \Ha_-^{\rm toy}\\ \Ha_+^{\rm toy} & 0
    \end{pmatrix}\,, \\
    \Ha_\pm^{\rm toy} &= \eta_4 \cos\f{\phi}{2} + \f{1}{2i}\sum_j \eta_j
\left[\U_j^\mp\T_j
-\T_j^\dag\U_j^{\pm}{}^\dag\right]\,,
  \end{aligned}
\end{equation}
where it is understood that all variables are now the toy-model
variables, e.g., $\cos\f{\phi}{2}=\diag(\f{\cos\phi_{\vec x}}{2})$. 
The dynamics of the spin phases $\phi_{\vec x}\in[-\pi,\pi)$ is
governed by the spin-model Hamiltonian
\begin{equation}
  \label{eq:sp_mod_ham_NC2}
  \beta H_{\rm noise} =  
  -\beta \sum_{\vec x,j}  \cos(\phi_{\vec x + \hat\jmath}-\phi_{\vec x})
  - h\sum_{\vec x}\cos(2\phi_{\vec x}) \,,
\end{equation}
as in Ref.~\cite{GKP2}. Here $\beta$ is the inverse temperature of the
spin model, and $h$ is a coupling which breaks the U$(1)$ symmetry of
the first term down to $\mathbb{Z}_2$.  
The dynamics of the toy-model link variables $\U_j^{\pm}(\vec x)\in
{\rm SU}(2)$ is governed by the action
\begin{equation}
  \label{eq:toy2}
  \begin{aligned}
         S_u     &=
2\hat\beta\sum_{j,\vec x} 
\cos\f{\phi_{\vec x}}{2}\cos\f{\phi_{\vec x+\hat\jmath}}{2}
\Re\tr[  \U_j^+(\vec x)  \U_j^-(\vec x)^\dag] \\ &+
\sin\f{\phi_{\vec x}}{2}\sin\f{\phi_{\vec x+\hat\jmath}}{2}
\Re\tr [ \U_j^+(\vec x)\sigma_3  \U_j^-(\vec x)^\dag \sigma_3]
 \,,
  \end{aligned}
\end{equation}
where $\hat\beta$ plays the role of gauge coupling. 
Expectation values are defined as follows: 
\begin{equation}
  \label{eq:toy_Ham_5}
  \la {\cal O} \ra = \f{\int D\phi \,e^{-\beta H_{\rm noise}[\phi]} \left[\f{\int
        D\U \, e^{-S_u[\phi,\U]}{\cal O}[\phi,\U]}{\int D\U\,
        e^{-S_u[\phi,\U]}}\right]}{\int D\phi \,e^{-\beta H_{\rm noise}[\phi]}} \,,
\end{equation}
where we have denoted $\int D\phi=\prod_{\vec x}
\int_{-\pi}^{+\pi}d\phi_{\vec x}$ and $D\U 
=\prod_{\vec x,j} d\U_j^+(\vec x)d\U_j^-(\vec x)$, with
$d\U_j^\pm(\vec x)$ the Haar measure. Notice the absence of
backreaction of the gauge links on the spins.
In practice, configurations are obtained by first sampling the spin
configurations $\{\phi_{\vec x}\}$ according to their Boltzmann weight
$e^{-\beta H_{\rm noise}[\phi]}$, and then, for a given $\{\phi_{\vec
  x}\}$, by sampling the spatial link configurations
$\{\U^{\pm}_j(\vec x)\}$ according to their Boltzmann weight
$e^{-S_u[\phi,\U]}$. 

The features that have been stripped from QCD in order to build the
toy model are those deemed irrelevant for the qualitative behavior of
eigenvalues and eigenvectors of the Dirac operator. What has been
kept is the presence of order in the configuration of the variables
governing the diagonal noise of the Hamiltonian, and the correlations
that such order induces on the spatial links. 
Due to our drastic simplifications [especially the decoupling of the
spin/Polyakov-line dynamics from that of the spatial links, see
Eq.~\eqref{eq:toy_Ham_5}] we do not expect any quantitative
correspondence between our model and lattice QCD, but just a
qualitative one. More precisely, there is no simple way to set the
parameters of the toy model to get quantitative agreement with
lattice QCD. In particular, intuition from QCD about scales (lattice
spacing, localization lengths\ldots) cannot be  used in the toy model,
as this has its own dynamics that set these scales. One might also be
worried by our choice $N_T=2$, which is known to be problematic in
QCD, and not likely to lead to good quantitative results
there. Nevertheless, this is a legitimate (and indeed the simplest)
choice one can make to build a toy model which qualitatively resembles
QCD with staggered fermions (see Ref.~\cite{GKP2} for a more detailed
discussion). In particular, one need not be worried 
about the fact that a very coarse lattice is needed in lattice QCD with
$N_T=2$ to reach the transition temperature: having decoupled the spin
dynamics from the rest, whether or not the spin system undergoes a
transition is entirely independent of $N_T$. The results of
Ref.~\cite{GKP2} show that the toy model described above is indeed
capable of reproducing the important features of the spectrum and of
the eigenmodes, both in the ordered and in the disordered phase.

\section{Numerical results}
\label{sec:num}

\begin{figure}[t]
  \centering
  \includegraphics[width=0.48\textwidth]{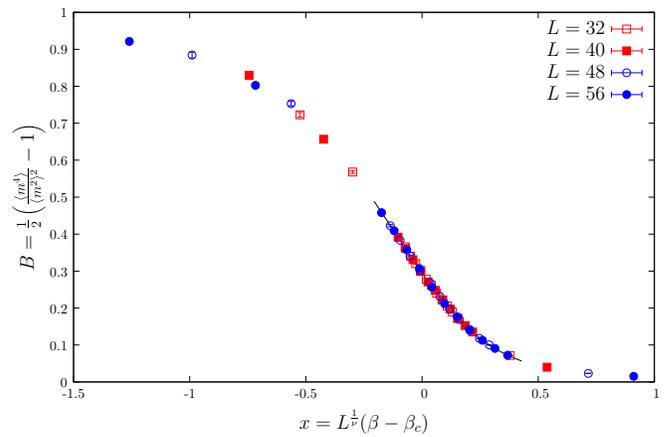}
  \caption{Finite-size-scaling analysis for the Binder cumulant $B$.
    The scaling function is also shown (solid line).}
  \label{fig:fss}
\end{figure}

In this section we report the results of a numerical study of the toy
model defined by Eqs.~\eqref{eq:toy1}--\eqref{eq:toy_Ham_5} in the
vicinity of the phase transition in the underlying spin model. Numerical
simulations near a critical point are hampered by critical slowing
down, but this problem can be overcome using a suitable cluster
algorithm. This is discussed in subsection \ref{sec:num1}, where we
report the results of a detailed finite-size-scaling study of the spin
model Eq.~\eqref{eq:sp_mod_ham_NC2}, aimed at determining the critical 
coupling and the universality class of the transition. 

We then proceed to study in our toy model the issues of localization
and chiral transition, the latter understood here as a singularity in
the spectral density at the origin.
The most effective
observables in pinning down the coupling(s) at which localization
appears and/or where a chiral transition takes place, 
are respectively the participation ratio of the lowest eigenmode and
the corresponding level spacing. This is discussed in subsection
\ref{sec:num2}, where we also report the results of our numerical study.

\subsection{Finite-size-scaling study of the spin model}
\label{sec:num1}

\begin{figure}[t]
  \centering
  \includegraphics[width=0.48\textwidth]{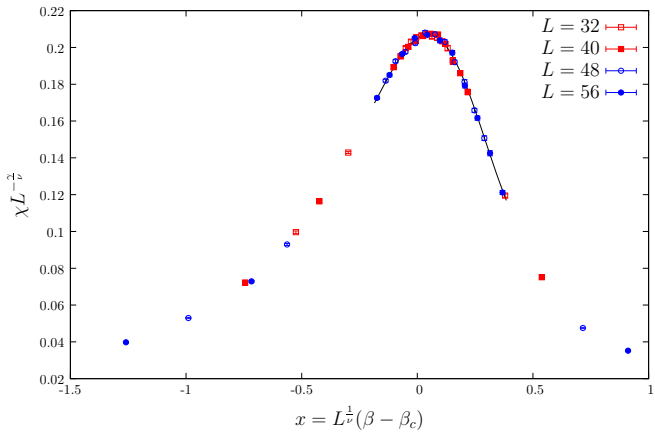}
  \caption{Finite-size-scaling analysis for the
    magnetic susceptibility $\chi$. The scaling function is also shown
    (solid line).} 
  \label{fig:fss2}
\end{figure}

We begin by studying the spin model on its own near the critical
point. The formulation of a cluster algorithm to this end is made
easier by noticing that Eq.~\eqref{eq:sp_mod_ham_NC2} can be recast as 
\begin{equation}
  \label{eq:spin1}
  \beta H_{\rm noise} =  
  -\beta \sum_{\vec x,j} \hat n_{\vec x}\cdot \hat n_{\vec x + \hat\jmath}
  - h\sum_{\vec x}\left[2\left(\hat n_{\vec x}\cdot \hat n_*\right)^2 -1\right]
 \,,
\end{equation}
where
\begin{equation}
  \label{eq:spin2}
  \hat n_{\vec x} =
  \begin{pmatrix}
    \cos\phi_{\vec x} \\ \sin\phi_{\vec x}
  \end{pmatrix}
\,,\qquad
  \hat n_* =
  \begin{pmatrix}
    1 \\ 0
  \end{pmatrix}\,.
\end{equation}
Near the transition $\hat n_{\vec x}$ tends to be aligned to $\pm\hat
n_*$, forming large clusters of like-oriented spins, and this leads to
long autocorrelation times in the simulation history. To overcome 
critical slowing down we thus employed a Wolff-type cluster
algorithm~\cite{Wolff:1988uh} consisting of the following steps.
\begin{enumerate}
\item Given a spin configuration, we pick a site at random and build a
  cluster, adjoining nearby sites $\vec x$ and $\vec x\pm\hat\jmath$ with
  probability 
  \begin{equation}
    \label{eq:spin3}
    \begin{aligned}
   P(\vec x,\vec x\pm\hat\jmath)&= 1 -e^{-2\beta |\hat n_{\vec x}\cdot\hat
     n_*| |\hat n_{\vec 
      x\pm\hat\jmath}\cdot\hat n_*|}\\ &=
     1 -e^{-2\beta |\cos\phi_{\vec x}| |\cos\phi_{\vec x+\hat\jmath}|}\,.
    \end{aligned}
  \end{equation}
\item Once the cluster is built, we flip $\hat n_{\vec x}\to -\hat
  n_{\vec x}$, i.e., we send $\phi_{\vec x}\to \pi\,\sgn(\phi_{\vec
    x})-\phi_{\vec x}$, for all sites $\vec x$ in the cluster.
\end{enumerate}
This algorithm is easily shown to respect detailed balance, but it
obviously fails at being ergodic. For this reason, we paired it with a
standard Metropolis algorithm, which restores ergodicity.

\begin{table}[t]
  \centering
  \begin{tabular}{l|l|l}
    \mbox{} & Our model & 3D Ising model \\
    \hline
    $\beta_c$ & 0.3023210(38) & --- \\
    $B^*$ & 0.2952(13) & 0.3022(13) \\
    $\nu$ & 0.6393(84) & 0.6301(8) \\
    $\gamma$ & 1.2660(26) & 1.237(2)
  \end{tabular}
  \caption{Critical point, critical Binder cumulant and critical
    exponents of our spin model, and of the 3D Ising model~\cite{blote}.}
  \label{tab:1}
\end{table}

We studied the model as a function of $\beta$ keeping the
symmetry-breaking term fixed at $h=1.0$. Defining the magnetization of 
the system as   
\begin{equation}
  \label{eq:spin4}
  m = \sum_{\vec x} \Re s_{\vec x} = \sum_{\vec x} \cos \phi_{\vec x}\,,
\end{equation}
we measured the susceptibility and the fourth-order Binder cumulant:
\begin{equation}
  \label{eq:spin5}
  \chi = \f{1}{V}\left( \la m^2\ra -\la m\ra^2\right)\,,\qquad
B = \f{1}{2}\left(\f{\la m^4\ra}{\la m^2\ra^2}-1\right)\,.
\end{equation}
Our definition of $B$ is such that $B\to 1$ in the disordered phase
and $B\to 0$ in the ordered phase. Near the critical point, $\beta_c$,
the expected behavior of $B$ and $\chi$ is
\begin{equation}
  \label{eq:fss_eq}
  \begin{aligned}
    B(\beta) &\approx f\left(L^{\f{1}{\nu}}(\beta-\beta_c)\right)\,, \\
    \chi(\beta) &\approx
    L^{\f{\gamma}{\nu}}g\left(L^{\f{1}{\nu}}(\beta-\beta_c)\right)\,. 
  \end{aligned}
\end{equation}
We thus fitted the numerical data in the range $\beta\in[0.302,0.303]$
and for the available volumes with the functional forms of
Eq.~\eqref{eq:fss_eq}, using polynomial approximations of $f$ and $g$
of increasing order, and assessing the error by means of constrained
fit techniques~\cite{Lepage:2001ym}. Our results for the critical
point, the critical 
exponents $\nu$ and $\gamma$, and the critical Binder cumulant $B^*$
are reported in Tab.~\ref{tab:1}. 
These values give an excellent ``collapse'' of the data points on a
single, volume-independent curve, as shown in Figs.~\ref{fig:fss} and
\ref{fig:fss2}.  
For comparison, in Tab.~\ref{tab:1} we report also the results of
Bl\"ote {\it et al.\/} for the 3D Ising model~\cite{blote}. The
tension in the results for $B^*$ and $\gamma$ is 
probably due to the 
fact that we are not including the effect of irrelevant couplings in
our analysis. Nevertheless, our results strongly support the fact that
the transition observed in our model belongs to the 3D Ising
universality class.

\subsection{Onset of localization and chiral transition in the toy model}
\label{sec:num2}

\begin{figure}[t]
  \centering
  \includegraphics[width=0.48\textwidth]{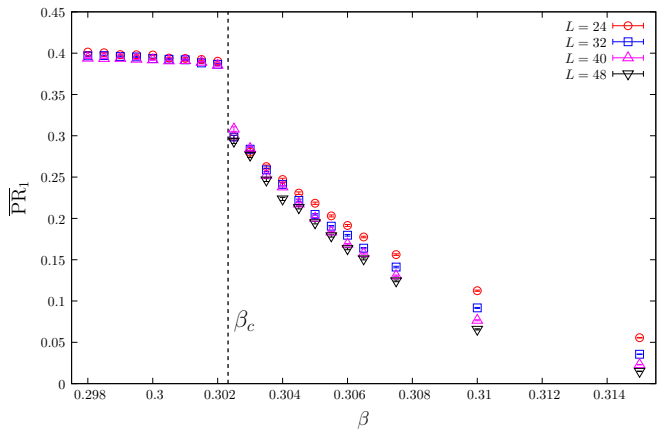}
  \caption{Average participation ratio of the first eigenmode as a
    function of $\beta$.}
  \label{fig:pr1}
\end{figure}

Let us discuss first the issue of localization. 
The simplest way to check for localization is to compute
the so-called ``participation ratio'', ${\rm PR}_n$, of the $n$th
eigenmode, $\psi_n$, defined as
\begin{equation}
  \label{eq:pr_def}
  {\rm PR}_n = \f{1}{V}{\rm IPR}_n^{-1} = \f{1}{V} \left[\sum_{\vec x}|
    \psi_n^\dag(\vec x) 
  \psi_n(\vec x)|^2\right]^{-1}\,,
\end{equation}
where ${\rm IPR}$ stands for ``inverse participation ratio'', and
$\psi_n^\dag \psi_n =  \sum_{a,k} (\psi_n)_{ak}^* (\psi_n)_{ak}$ stands
for summation over the color and temporal-momentum degrees of
freedom. Here $V=L^3$ is the spatial volume. 
If the $n$th mode is localized, then the average of ${\rm PR}_n$ over
configurations, which we denote by $\overline{\rm PR}_n$, is expected
to vanish in the large-volume limit.
On the other hand, for delocalized modes this quantity becomes
constant at large volume. We already know from Ref.~\cite{GKP2} that
localized modes appear first near the origin, so in order to check whether
there are localized modes or not, it is sufficient to compute the
participation ratio of the first eigenmode, and check how it changes
with the volume. In Fig.~\ref{fig:pr1}
we show the average participation ratio of the first eigenmode,
$\overline{{\rm PR}}_1$, as a function of $\beta$ for different
system sizes, namely $L=24,32,40$ and, in the ordered phase only, also
$L=48$. The localization properties of the lowest mode are clear
below $\beta_c$ and well above it. In the disordered phase the lowest
mode is delocalized, while it is localized deep in the ordered
phase. Starting from large $\beta$ and going down towards $\beta_c$
the scaling with $V$ becomes slower, and very close to $\beta_c$
the participation ratio actually grows up to $L=40$.
Nevertheless, $\overline{{\rm PR}}_1$
displays a jump at $\beta_c$, and the largest volume always gives 
the smallest participation ratio. We take these findings as an
indication that also right above $\beta_c$ the lowest eigenmode has
the tendency to localize. This tendency is, however, hampered by the
fact that the typical localization length is bigger than or comparable to
the system sizes under consideration. As a consequence, the would-be
(lowest) localized eigenmode is effectively delocalized on the whole
lattice, thus having a strong overlap with the extended modes, and
therefore mixing easily with them under fluctuations of the spins and
of the gauge fields. Moreover, we expect its participation ratio
to grow until the system is big enough to accommodate a localized mode,
whereas it will start to decrease for even larger sizes. 
In conclusion, we expect that for sufficiently
large systems the lowest eigenmode is localized as soon as
$\beta>\beta_c$. The closer one is  to $\beta_c$, the larger the
system has to be for localization to be fully visible.

\begin{figure}[t]
  \centering
  \includegraphics[width=0.48\textwidth]{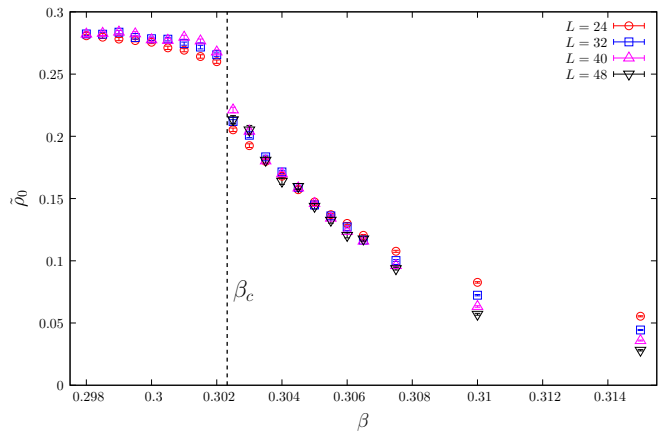}
  \caption{The quantity $\tilde\rho_0$, defined in
    Eq.~\protect\eqref{eq:unf_def3_bis}.}
  \label{fig:dellam1}
\end{figure}

\begin{figure}[t]
  \centering
  \includegraphics[width=0.48\textwidth]{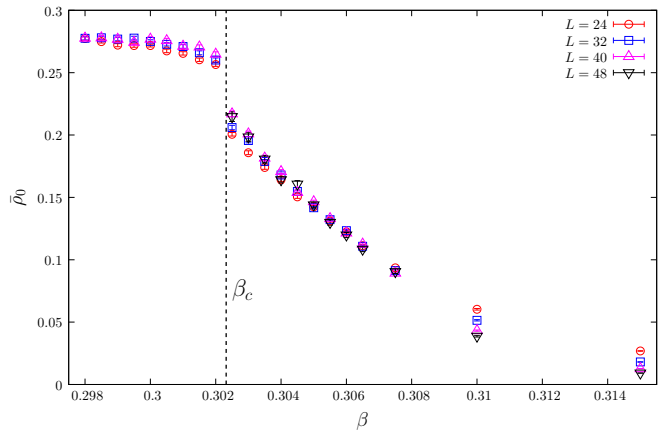}
  \caption{The quantity $\bar\rho_0$ defined in
    Eq.~\protect\eqref{eq:barrho_def_bis}.} 
  \label{fig:lam1}
\end{figure}

Let us consider next the issue of the chiral transition.
In principle (and by definition), this issue should be studied by
analyzing the spectral density near the origin. In practice, however,
this is very hard in the vicinity of the critical point, and could be
done reliably only using high statistics and large volumes, in order
to sample properly the near-zero spectral region. Rather than
attempting a (difficult) direct measurement, we relied upon the
following relation:
\begin{equation}
  \label{eq:unf_def3_bis}
  \tilde\rho_0 \equiv \f{1}{V\la \lambda_2 -\lambda_1\ra}
  \mathop\to_{V\to\infty} \f{\rho(0)}{V}\,,
\end{equation}
which is based on the following argument. In the large-volume limit,
the spectral density at the origin is equal to the inverse of the
average level spacing in the near-zero spectral region. In the same
limit, and for fixed $j$, one has for the
eigenvalues of the Dirac operator (and thus for those of our toy-model
Hamiltonian) that $\lambda_j\to 0$. Eq.~\eqref{eq:unf_def3_bis} then
follows. This applies to any fixed $j$, but of course one expects that
for too large $j$ the finite-size effects would completely obscure the
limit (however, see below for some numerical results for $j=2,3$). 
In Fig.~\ref{fig:dellam1} we show $\tilde\rho_0$ as a function of
$\beta$ for the available system sizes. It is clear that below
$\beta_c$ this quantity tends to a 
finite constant as the volume is increased. For our largest values of
$\beta$ above $\beta_c$, on
the contrary, there is a clear tendency for $\tilde\rho_0$ to
vanish as $V\to\infty$. The region which is most difficult to
understand is right above $\beta_c$. There $\tilde\rho_0$ apparently
tends to a finite constant, different from the one right below
$\beta_c$. Although it is possible that there are two jumps in
$\tilde\rho_0$, one at $\beta_c$ and another at some higher value of
$\beta$ where $\tilde\rho_0$ jumps to zero, we think that there is a
more plausible explanation for this behavior. In fact, as we have
already mentioned above, the relative smallness of the system, which
causes the lowest mode to be effectively delocalized, is also
responsible for its mixing with nearby modes under fluctuations of
spins and link variables. The behavior of the lowest mode is thus
expected to be similar in all respects to what is found in the
disordered phase, and more generally the low end of the spectrum is
expected to look the same as it looks in the disordered phase. This
includes a nonzero spectral density near the origin. It is likely that
for large enough systems, $\tilde\rho_0$ will start to show a
nontrivial scaling with $V$, indicating the vanishing of the spectral
density at the origin in the thermodynamic limit. In any case, whether
$\rho(0)$ vanishes right above $\beta_c$ or not, it is clear that at
$\beta_c$ it displays a singularity. This indicates that the system
has a chiral transition at $\beta_c$.


\begin{figure}[t]
  \centering
  \includegraphics[width=0.48\textwidth]{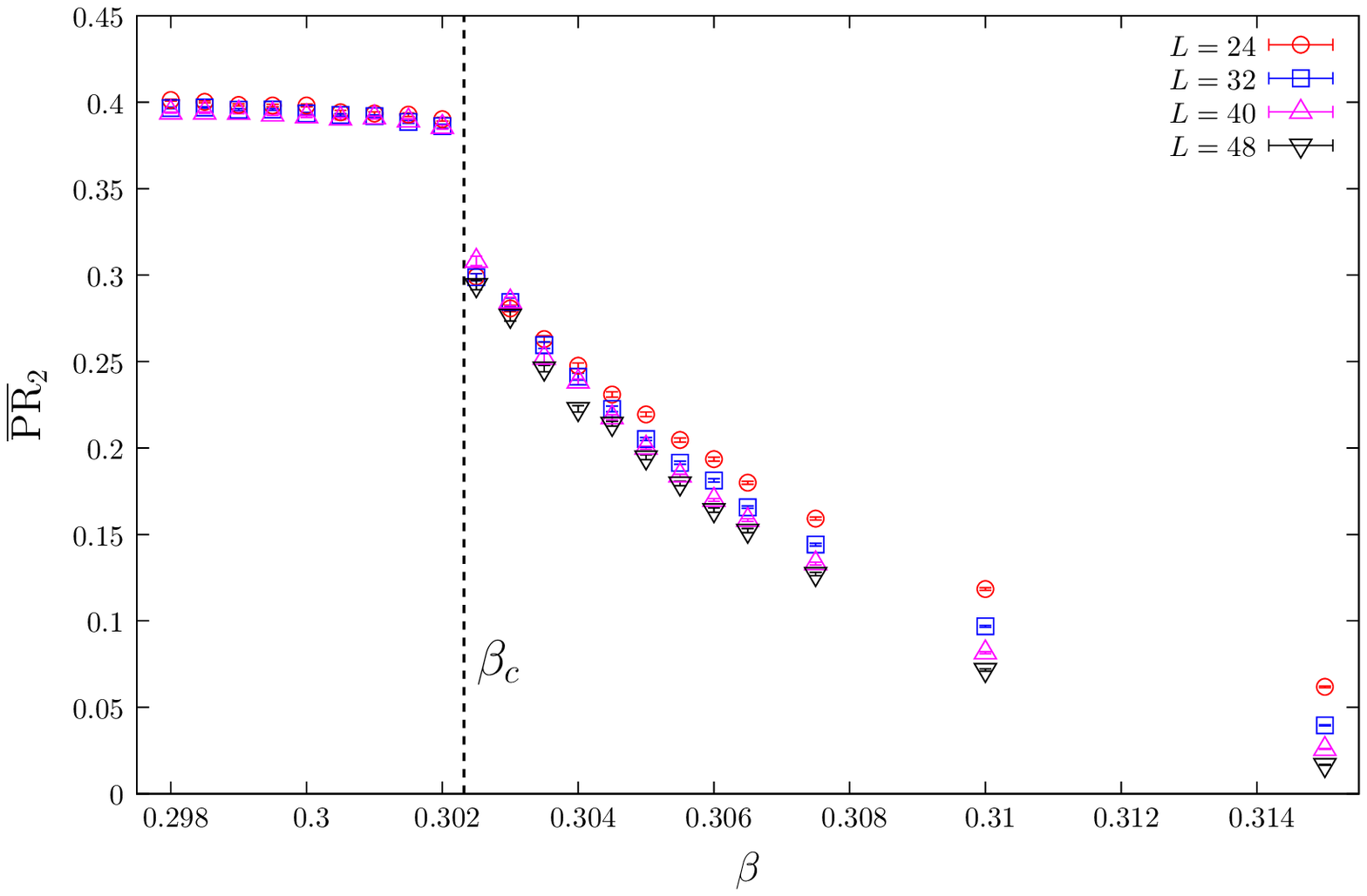}\hfil
  \includegraphics[width=0.48\textwidth]{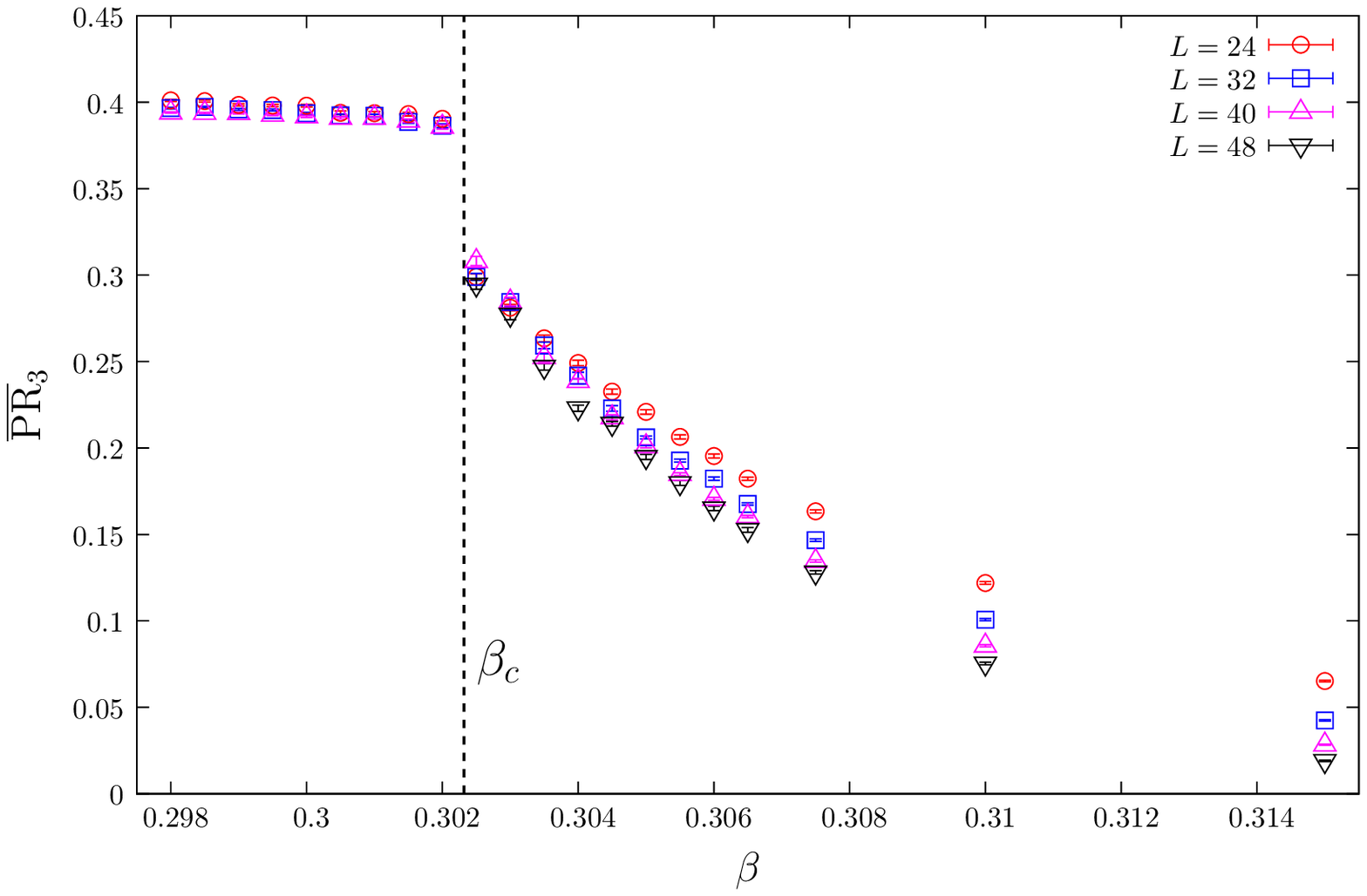}
  \caption{Average participation ratio of the second and third eigenmode as a
    function of $\beta$.}
  \label{fig:pr23}
\end{figure}

An alternative way of determining $\rho(0)$ is based on its relation
with the expectation value of the lowest eigenvalue,
$\la\lambda_1\ra$. In the disordered phase, where $\rho(0)\neq 0$, the
probability distribution of the lowest eigenvalue is expected to be
described by the appropriate ensemble of chRMT. In the case at hand,
this should be the symplectic ensemble for the quenched theory in the
trivial topological sector, and so~\cite{Forrester,Kaneko,
BerbenniBitsch:1998sy,Nishigaki:1998is,Nishigaki:2016nka} 
\begin{equation}
  \label{eq:chGSE_l1_bis}
  p_1(z) = \sqrt{\f{\pi}{2}} z^{\f{3}{2}} I_{\f{3}{2}}(z) e^{-\f{z^2}{2}}\,,
\end{equation}
where $z = \lambda_1 \pi \rho(0)$. From this one obtains the appropriate
proportionality factor between $\la\lambda_1\ra$ and $\rho(0)$, namely
\begin{equation}
  \label{eq:chGSE_l1_3_bis}
\rho(0) = \sqrt{\f{e}{2\pi}} \f{1}{\la \lambda_1 \ra}\,.
\end{equation}
For localized modes one expects instead that the corresponding
eigenvalues obey Poisson statistics. In this case,
assuming a power-law behavior $\rho(\lambda)=CV\lambda^\alpha$ for
the spectral density near the origin, one finds~\cite{KGT} that $\la
\lambda_1 \ra \sim V^{-\f{1}{1+\alpha}}$, and in particular $\rho(0) =
\f{1}{\la \lambda_1 \ra}$ for $\alpha=0$. Our results for
$\bar\rho_0$,
\begin{equation}
  \label{eq:barrho_def_bis}
\bar\rho_0\equiv \sqrt{\f{e}{2\pi}}\f{1}{V\la \lambda_1\ra}\,,
\end{equation}
are shown in Fig.~\ref{fig:lam1}. Comparing this with
Fig.~\ref{fig:dellam1} we see that the chRMT result works well below
$\beta_c$, while it works less and less well
as $\beta$ increases above $\beta_c$. In particular, for large $\beta$
one has that $\bar\rho_0$ tends to vanish as the volume is increased,
signaling a vanishing spectral density at the origin. 
As before, the region right above $\beta_c$ is the one where things
are less clear. A nonvanishing $\rho(0)$ accompanied by localization
of the lowest modes right above $\beta_c$ should yield a $\bar\rho_0$
appreciably smaller than $\rho(0)$, and so of $\tilde\rho_0$, while
the two quantities compare well. This is most likely another
consequence of the smallness of the system size compared to what would
be required to properly investigate the region near the critical
point. In fact, the effective delocalization and easy mixing of the
lowest mode mentioned above leads to correlations building up among
eigenvalues, thus leading to a chRMT-like statistical behavior, which
should go over to Poisson behavior as the system size increases. 

\begin{figure}[t]
  \centering
  \includegraphics[width=0.48\textwidth]{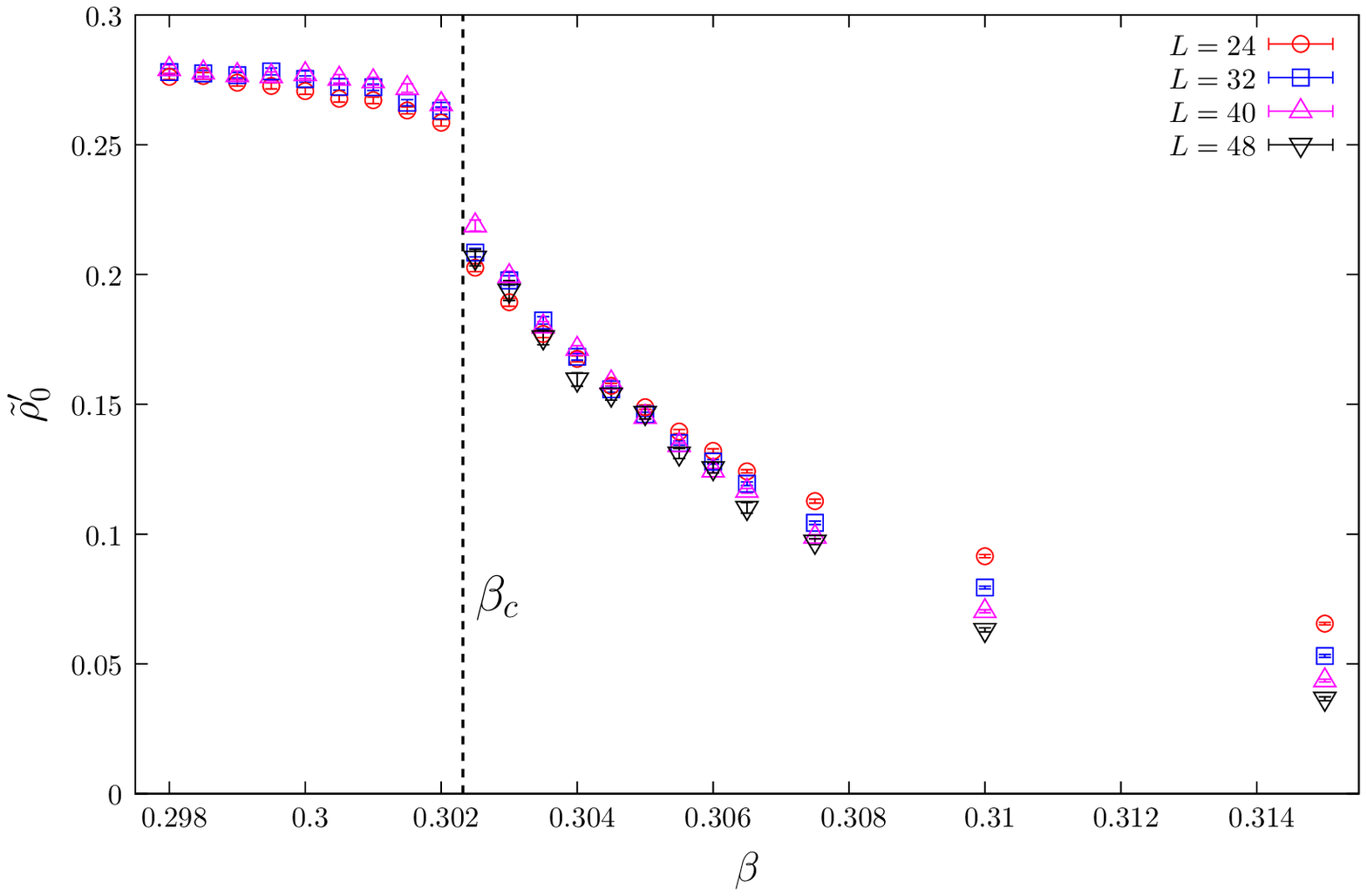}\hfil
  \includegraphics[width=0.48\textwidth]{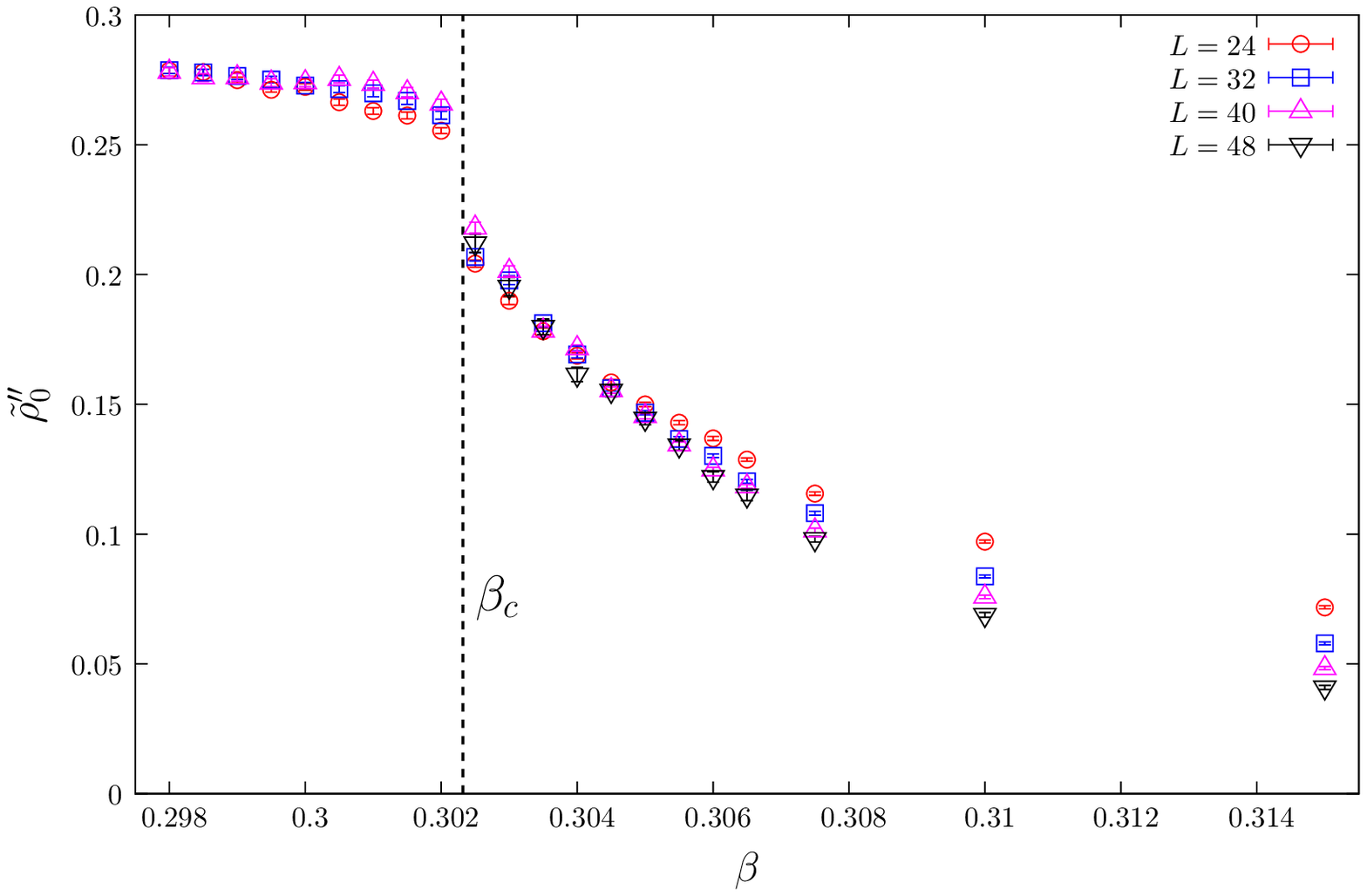}
  \caption{The quantities $\tilde\rho_0'$ and $\tilde\rho_0''$, defined in
    Eq.~\protect\eqref{eq:unf_def_higher}.}
  \label{fig:dellam23}
\end{figure}

For completeness, we conclude this section by showing our numerical
results concerning the second and third lowest eigenmodes. In
Fig.~\ref{fig:pr23} we show the average participation ratios
$\overline{{\rm PR}}_2$ and $\overline{{\rm PR}}_3$. The situation is
entirely analogous to that encountered when studying the lowest mode,
with similar finite-size effects near the transition which slow down the
localization of these modes. In Fig.~\ref{fig:dellam23} we show the
quantities
\begin{equation}
  \label{eq:unf_def_higher}
  \tilde\rho_0' \equiv \f{1}{V\la \lambda_3 -\lambda_2\ra}\,,\qquad
  \tilde\rho_0'' \equiv \f{1}{V\la \lambda_4 -\lambda_3\ra}\,,
\end{equation}
which in the large-volume limit should also approach $\f{\rho(0)}{V}$.
In this case the volume scaling is somewhat more clear, with the
tendency to go to zero as $V$ is increased showing up for lower values
of $\beta$. These results clearly do not change the conclusions
discussed above.

\section{Conclusions and outlook}
\label{sec:concl}

There are by now several hints at a close connection between the
deconfining and chiral transitions and localization of the lowest
eigenmodes of the Dirac operator.
In this paper we have further studied the toy model of
Ref.~\cite{GKP2}, which mimics the effects of the ordering of Polyakov
loops in QCD, i.e., deconfinement, on the spectral density of the low
Dirac eigenmodes and the corresponding localization properties. In
particular, we have focused on the region near the magnetic
transition of the underlying spin model, which corresponds to
deconfinement in a gauge theory. We have then studied numerically the
localization properties of the lowest eigenmode, and the spectral
density at the origin. Our findings are consistent with a chiral
transition taking place in correspondence with the magnetic
transition, accompanied by the appearance of localized modes. This
further supports our expectation that deconfinement plays a major role
in the chiral transition and in the localization of the low Dirac
modes observed in QCD.

There are, however, several aspects that deserve further
study. The presence of a chiral transition in our toy model
when the spins get ordered is quite clear, since the spectral density
at the origin shows a jump there. However, it is not clear yet if such
a jump is from the finite value of $\rho(0)$ in the disordered phase
to zero in the ordered phase, or to a different finite value. Although
the latter possibility seems unlikely, nevertheless the presence of
strong finite-size effects makes it difficult to extrapolate to the
infinite-volume limit. The origin of such effects lies in the fact
that although the lowest modes would like to be localized, their
typical localization length is bigger than the system sizes at our
disposal. This makes those modes effectively delocalized on our finite
lattices, and so easily mixed by fluctuations with other nearby
modes. In turn, this is probably responsible for a smaller typical
level spacing between the first two eigenvalues, from which the
spectral density was extracted. Consequently, we are probably
overestimating $\rho(0)$. Moreover, the lowest eigenmode correlates
with the nearby modes, which results in statistical properties closer
to those predicted by chRMT than to those expected for localized
modes, which should obey Poisson statistics. In order to overcome
these problems, and unveil the true nature of the lowest modes,
bigger lattices should be employed.

This situation should be contrasted to that found with unimproved
staggered fermions on coarse lattices~\cite{GKKP}. In that case the
coincidence of deconfinement, chiral transition and appearance of
localized modes is more clean cut. A possible explanation of the
difference lies in the different nature of the deconfining transition
in that system, which is a first-order transition, and the magnetic
transition in our toy model, which is a second-order phase transition
of the 3D Ising universality class. In the case at hand, the presence
of a huge correlation length near the critical point, and at the same
time the fact that the magnetization is very small there, makes it
more difficult for the low modes to properly localize. As we said
above, this is expected to be the source of the large finite-size
effects observed in our determination of $\rho(0)$.

Despite these difficulties, we think that our results confirm those of
previous studies in other models, in showing that deconfinement,
chiral transition and localization are closely tied to each
other. There are several possible extensions of the present study. One
obvious possibility is to consider our toy model for gauge group
SU$(3)$, thus making it closer to QCD. This involves a different spin
model to mimic the behavior of the Polyakov lines than the one
employed here (see Ref.~\cite{GKP2} for details). A more interesting
possibility is to extend the toy model to the case of adjoint
fermions: this could help in understanding why for adjoint fermions
deconfinement and chiral restoration take place at different
temperatures~\cite{adjoint}.

\section*{Acknowledgements}
This work is partly supported by OTKA under the grant OTKA-K-113034.
TGK is supported by the Hungarian Academy of Sciences under
``Lend\"ulet'' grant No. LP2011-011.

\end{document}